\documentclass[a4paper,11pt]{article}
\newcommand{\be} {\begin{equation}}
\newcommand{\ee} {\end{equation}}
\newcommand{\ben} {\begin{equation*}}
\newcommand{\een} {\end{equation*}}
\newcommand{\bdm} {\begin{displaymath}}
\newcommand{\edm} {\end{displaymath}}
\newcommand{\bc} {\begin{center}}
\newcommand{\ec} {\end{center}}
\newcommand{\bea} {\begin{eqnarray}}
\newcommand{\eea} {\end{eqnarray}}
\newcommand{\bean}{\begin{eqnarray*}}
\newcommand{\eean}{\end{eqnarray*}}

\newcommand{\bfig} {\begin{figure}}
\newcommand{\efig} {\end{figure}}
\newcommand{\btab} {\begin{tabular}}
\newcommand{\etab} {\end{tabular}}

\usepackage{epsfig}
\usepackage{amssymb}
\usepackage{amsmath}

\def\NPB{{\em Nucl. Phys.} B}
\def\PLB{{\em Phys. Lett.}  B}

\def\PRD{{\em Phys. Rev.} D}
\def\ZPC{{\em Z. Phys.} C}

\begin{document}

\bc
{\bf \Large A COMPREHENSIVE APPROACH TO STRUCTURE FUNCTIONS}
\footnote{Summary of talk given at the XIth International Conference on
Elastic and Diffractive Scattering, Blois 2005}\\
\vspace*{0.3in}

{\Large A. DONNACHIE }\\
\vspace{.1in}
{\it \large School of Physics and Astronomy, University of Manchester,}\\
{\it \large Manchester M13 9PL, England}
\ec
\vspace{.1in}
\bc
{\bf Abstract}
\ec
{\small A model is presented based on a dipole picture with a 
hard and a soft pomeron. It is assumed that large dipoles couple to the soft 
pomeron and small dipoles couple to the hard pomeron. Most of the parameters 
of the model are predetermined from proton-proton scattering and the only free 
parameter is the radius $R_c$, defining the transition from small to large 
dipoles. This is fixed by the proton structure function $F_2(x,Q^2)$. The 
model then successfully predicts $F_2^c(x,Q^2)$, $F_2^L(x,Q^2)$, $J/\psi$ 
photoproduction, $\gamma^* p \to \gamma p$, $\sigma_{\gamma p}^{\rm Tot}(s)$, 
$\sigma_{\gamma \gamma}^{\rm Tot}(s)$ and  $F_2^{\gamma}(x,Q^2)$.}
\vspace*{0.2in}

\section{Objective}

Deep inelastic scattering at small $x$ can be described successfully by a 
two-component model \cite{DL} comprising the soft non-perturbative pomeron of 
hadronic interactions, intercept $\sim 0.08 - 0.1$, and a hard pomeron, 
intercept $\sim 1.4$. Although phenomenologically successful, it does not 
explain, for example, the relative strengths of the hard and soft pomeron in 
deep inelastic scattering and in $J/\psi$ photoproduction, or why the charm 
structure function of the proton is dominated by the hard pomeron.

To answer these questions requires a specific model for the diffractive 
process, consideration of the particle wave functions and simultaneous 
treatment of several reactions to separate the dynamics of diffraction from 
wave-function effects. The intent is to obtain a global description of a 
variety of data in a simple model, not necessarily to obtain a detailed fit, 
although in practice the model is surprisingly successful despite its 
simplicity.

\section{The model}

The model proposed \cite{DD02} is a simple one, based on a dipole picture with 
two pomerons in which small dipoles couple to the hard pomeron and large 
dipoles couple to the soft pomeron. Hadron-hadron, photon-hadron and 
photon-photon reactions are treated in a uniform approach. This is based on 
the ``Heidelberg Model'' \cite{Nac} of soft diffraction and the proton is 
considered as a quark-diquark system, that is effectively as a dipole. The 
parameters of the photon and proton wave functions and the dipole-dipole cross 
section are taken from applications of that model to soft hadronic \cite{Heid1}
and photoproduction reactions \cite{Heid2}. These parameters remain unchanged 
throughout, the criteria for defining small and large dipoles being obtained 
from the proton structure function. All other processes are controlled by the 
relevant particle wave functions.

\begin{figure}[t]
\begin{center}
\begin{minipage}{10cm}
\epsfxsize10cm
\epsffile{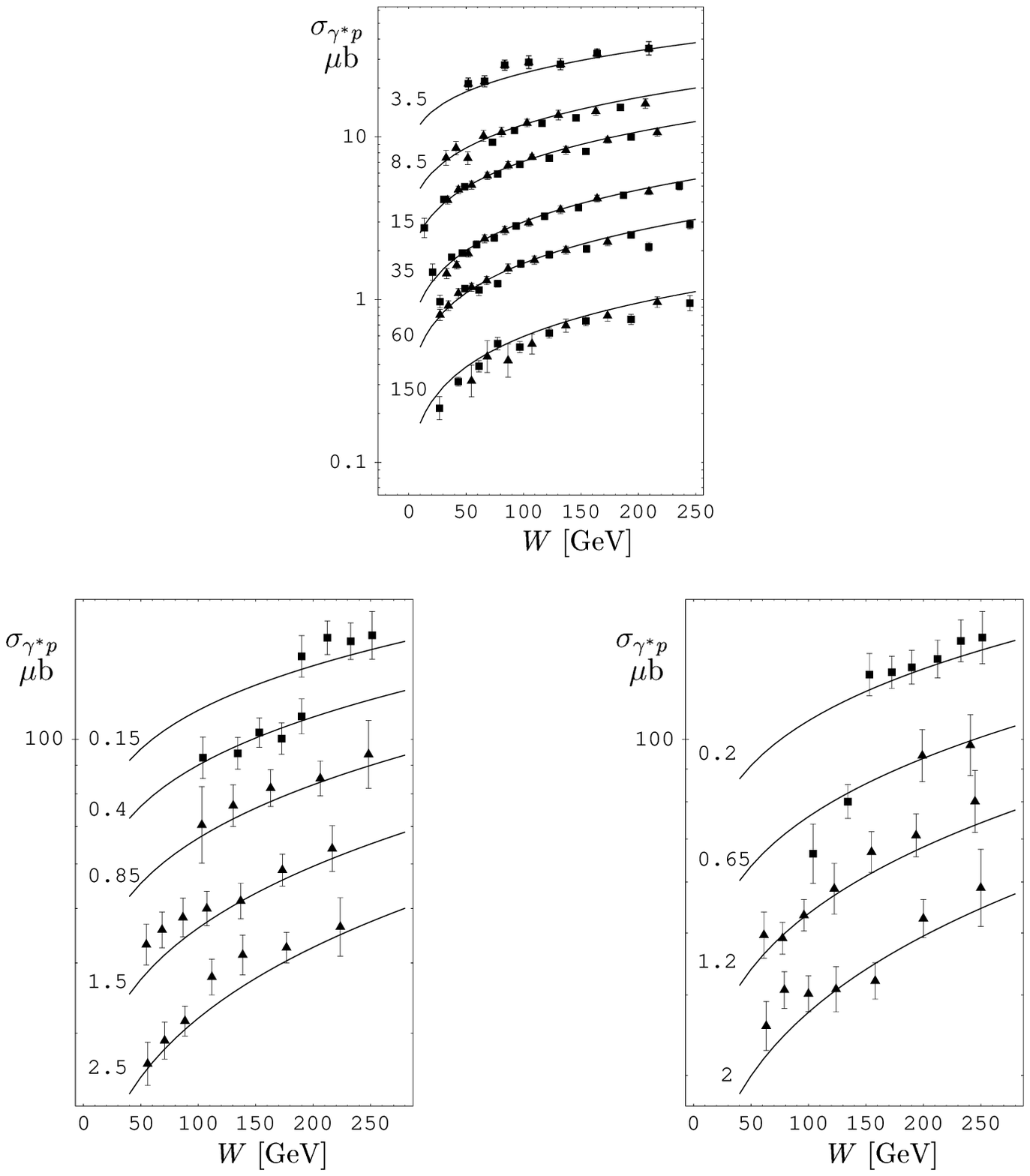}
\end{minipage}
\end{center}
\caption{Fit to $\sigma^{\rm Tot}_{\gamma^* p}(Q^2,W)$ for different values of 
$Q^2$.The data are from ZEUS \cite{ZEUS} (squares) and H1 \cite{H1} 
(triangles).}
\end{figure}

The standard perturbative expressions were used for the photon wave functions 
with a $Q^2$-dependent quark mass $m_{eff}(Q^2)$ for photons of low virtuality 
(including real).
\bea
m_{eff}(Q^2)  = & m_f &+m_{0q}(1-Q^2/Q_0^2)~~~~Q^2 \leq Q_0^2,\nonumber\\
                & m_f &~~~~~~~~~~~~~~~~~~~~~~~~~~~~~Q^2 \geq Q_0^2.
\eea
The masses used were $m_{0q} = 0.19$ GeV for the light quarks, 0.31 GeV for 
the strange quark and 1.25 GeV for the charm quark.

A Gaussian wave function was assumed for the proton:
\be
\psi_p(\vec R) = \frac{1}{\sqrt{2\pi}}exp\Big(-\frac{R^2}{4R_p^2}\Big).
\ee
The radius $R_p$ was chosen to reproduce the logarithmic slope of elastic 
$p p$ scattering at $\sqrt{s} = 20$ GeV, giving $R_p = 0.75$ fm. The $J/\psi$ 
wave function is taken as that of a massive vector current with mass $m_c$ and 
with the radial dependence modelled by a Gaussian, the radius parameter being 
fixed by the electromagnetic decay width. Full details can be found in 
\cite{Heid2}.

The Heidelberg Model is defined at $W =\sqrt{s} = 20$ GeV, so energy dependence
has to be introduced by hand by dividing the amplitude into a soft and a hard 
part:
\be
T_{ab \to cd} = iW^2\Big(T^{\rm soft}_{ab \to cd}(W/W_0)^{2\epsilon_s}+
T^{\rm hard}_{ab \to cd}(W/W_0)^{2\epsilon_h}\Big)
\ee
with $W_0 = 20$ GeV, $\epsilon_s = 0.08$ and $\epsilon_h = 1.42$.

To be economical with parameters a sharp cut was introduced, assuming that only
the soft pomeron couples if both dipoles are larger than a certain value $R_c$ 
whereas the hard pomeron couples if at least one of the dipoles is smaller than
$R_c$.

\begin{figure}[t]
\begin{center}
\begin{minipage}{12cm}
\epsfxsize12cm
\epsffile{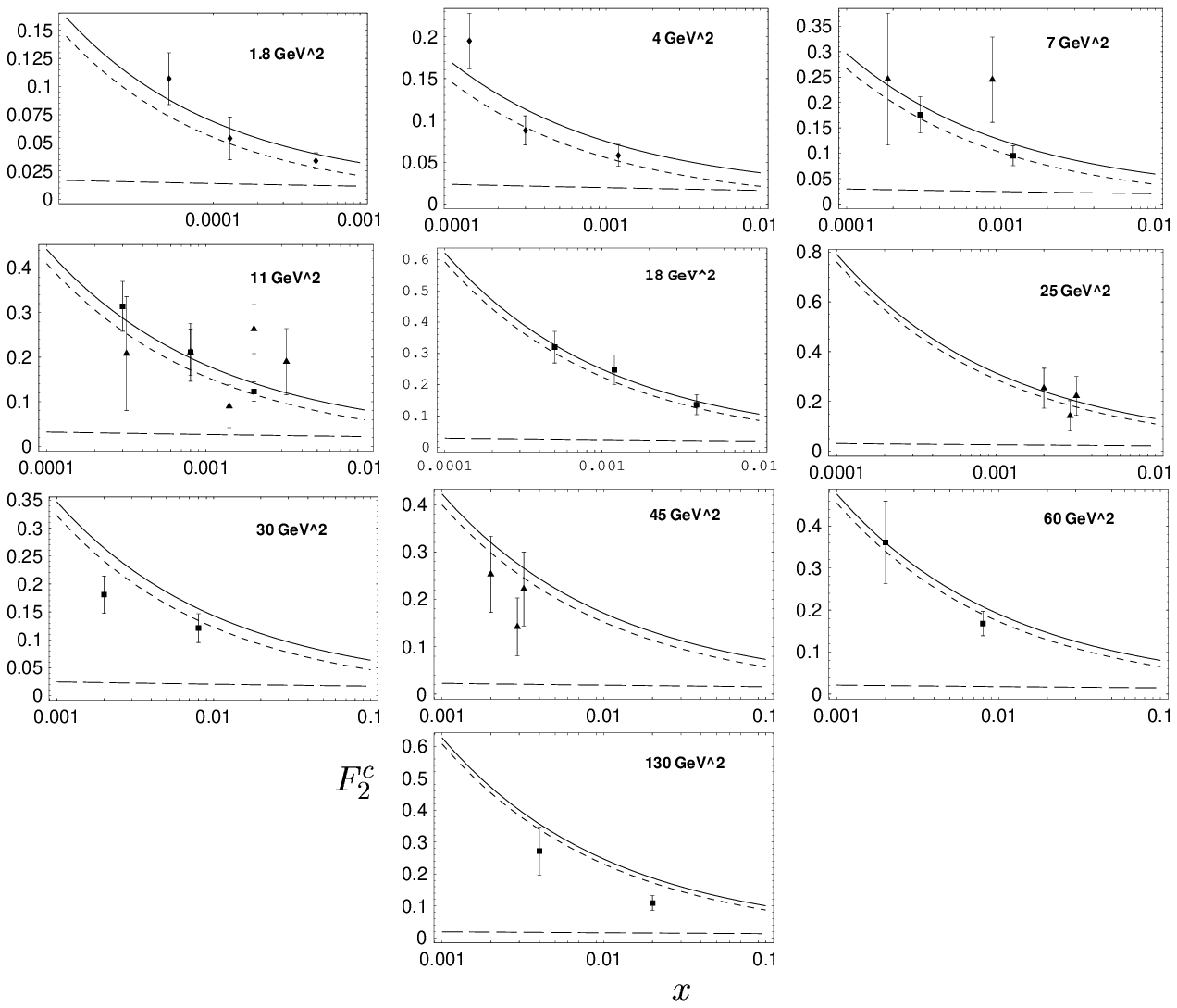}
\end{minipage}
\end{center}
\caption{The charm contribution to the proton structure function $F_2^c$ for 
different values of $Q^2$ as indicate in the figures. The solid line is the
full result, the short-dashed line the hard pomeron contribution and the 
long-dashed line the soft pomeron contribution. The data are from ZEUS 
\cite{ZEUSc} (squares) and H1 \cite{H1c} (triangles).}
\end{figure}

\section{$\gamma^* p$ reactions}

The parameter $R_c$, the only free parameter in the model, was obtained by 
fitting the proton structure function, or rather the total $\gamma^* p$ cross 
section, from $Q^2 \approx 0$ to 150 GeV$^2$. The result is $R_c = 0.22$ fm 
and examples of the fit are shown in figure 1.

With $R_c$ fixed, $F_2^c(x,Q^2)$ and $F_L(x,Q^2)$ can be predicted and 
agreement with the corresponding data is excellent. The model is compared 
with the $F_2^c$ data in figure 2. In both $F_2^c$ and $F_2^L$ the photon wave 
function is concentrated at smaller distances than in $F_2$, so the hard 
pomeron is already dominant at moderate energies. Although the increase of the 
hard-pomeron contribution in $F_L$ with increasing $Q^2$ is not quite as strong
as in $F_2$ the soft-pomeron contribution is even more suppressed so the hard 
pomeron is dominant sooner in $F_L$ than in $F_2$. We emphasize that these are 
purely wave-function effects. The model provides an explanation for the 
phenomenological result \cite{DLcharm} that $F_2^c(x,Q^2)$ is completely 
dominated by the hard pomeron, as can be seen in figure 2.

The ratio of the soft pomeron to the hard pomeron in $J/\psi$ photoproduction 
is much larger than in $F_2^c$ at comparable energies. This is again a 
wave-function effect. The virtual $c\bar c$ pair in the photon wave function 
has an extension $\approx 1/m_c$. The $J/\psi$ has a much larger radius, so 
the overlap of the charm part of the photon wave function with the $J/\psi$ 
wave function obtains a larger contribution from $R > R_c$ than does the 
square of the charm part of the photon wave function. The model describes
the $J/\psi$ total cross section well \cite{DD02}, either assuming a constant 
logarithmic slope of $b = 6$ GeV$^{-2}$ or with a slope varying with energy 
as predicted by Regge theory. For the latter the slope of the soft-pomeron 
trajectory was taken as the standard $\alpha_{P_s} = 0.25$ GeV$^{-2}$ and 
that of the hard-pomeron trajectory as $\alpha_{P_h} = 0.1$ GeV$^{-2}$ with
$b = 6$ GeV$^{-2}$ at $\sqrt{s} = 20$ GeV.

It is straightforward to predict the $\gamma p$ total cross section and the 
$\gamma^* p \to \gamma p$ (DVCS) data, both the $Q^2$ dependence at fixed 
$\sqrt{s}$ and the energy dependence at fixed $Q^2$. All three predictions are 
in excellent agreement \cite{DD02} with data. The model predicts a significant
contribution from the hard pomeron to the photoproduction cross section,
similar to that found in \cite{DL}. However the data do not demand a
contribution from the hard pomeron due to the comparatively large 
experimental errors at high energy.

\begin{figure}[t]
\begin{center}
\begin{minipage}{10cm}
\epsfxsize10cm
\epsffile{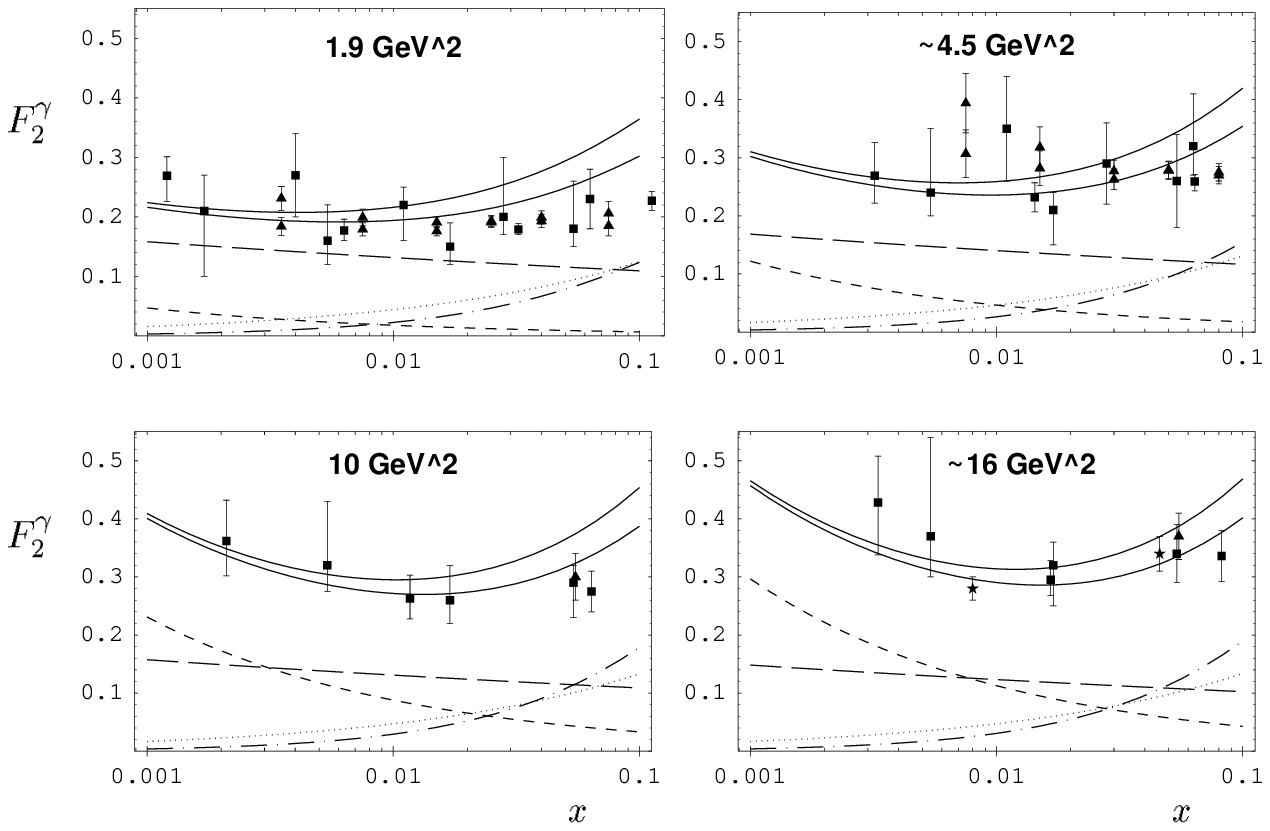}
\end{minipage}
\end{center}
\caption{The photon structure function $F_2^\gamma(x,Q^2)/\alpha$ for different
values of $Q^2$. The separate contributions are: soft pomeron (long dashes),
hard pomeron (short dashes), box diagram (dot-dashes), Regge (dots). The two 
solid lines are the combined results including the uncertainty in the Regge 
term. The data are from OPAL \cite{OPALa} (boxes), L3 \cite{L3a} (triangles)
and ALEPH \cite{ALEPHa} (stars).}
\end{figure}

\section{$\gamma^{(*)}\gamma^{(*)}$ reactions}

With the same approach and the same parameters the $\gamma\gamma$, $\gamma^*
\gamma$ ($F_2^\gamma$) and $\gamma^*\gamma^*$ cross sections can be obtained. 
As some of the data are at relatively low energies the Reggeon contribution 
and the box diagram (the quark-parton-model contribution) have to be included. 
The Regge contribution was estimated using the form given in \cite{DDR00} and 
the analytical results of \cite{Bud75} were used for the box diagram. The 
principal difference between $\gamma\gamma$ and $\gamma p$ at high energies 
comes from the singularity of the photon wave function at the origin, which 
favours the hard-pomeron component. This is apparent even for the scattering 
of real photons. The energy dependence of the $\gamma\gamma$ total cross 
section is not compatible with the soft pomeron alone, and the additional 
contribution of the hard pomeron is required. Once again the predictions of 
the model are in excellent agreement \cite{DD02} with the data. The model 
predictions for $F_2^\gamma/\alpha$ are equally satisfactory. The comparison 
with data is made in figure 3. Agreement with experiment is good 
for small $x$, but at large $x$ the increasing importance of the Regge term
leads to an increasing uncertainty in the predictions. Nonetheless they remain 
satisfactory for $x < 0.1$. The predictions for $\sigma^{\gamma^*\gamma^*}$ are
in fair agreement with the limited data. The only reactions for which the model
does not provide a good description of the data are $\gamma^{(*)}\gamma \to 
c\bar{c}X$, the predictions being about a factor of 2 lower than the L3 
\cite{L3} cross section for real photons and the OPAL result \cite{OPAL} for 
the charm contribution to the photon structure function at small $x$.

\begin{figure}[t]
\begin{center}
\begin{minipage}{10cm}
\epsfxsize10cm
\epsffile{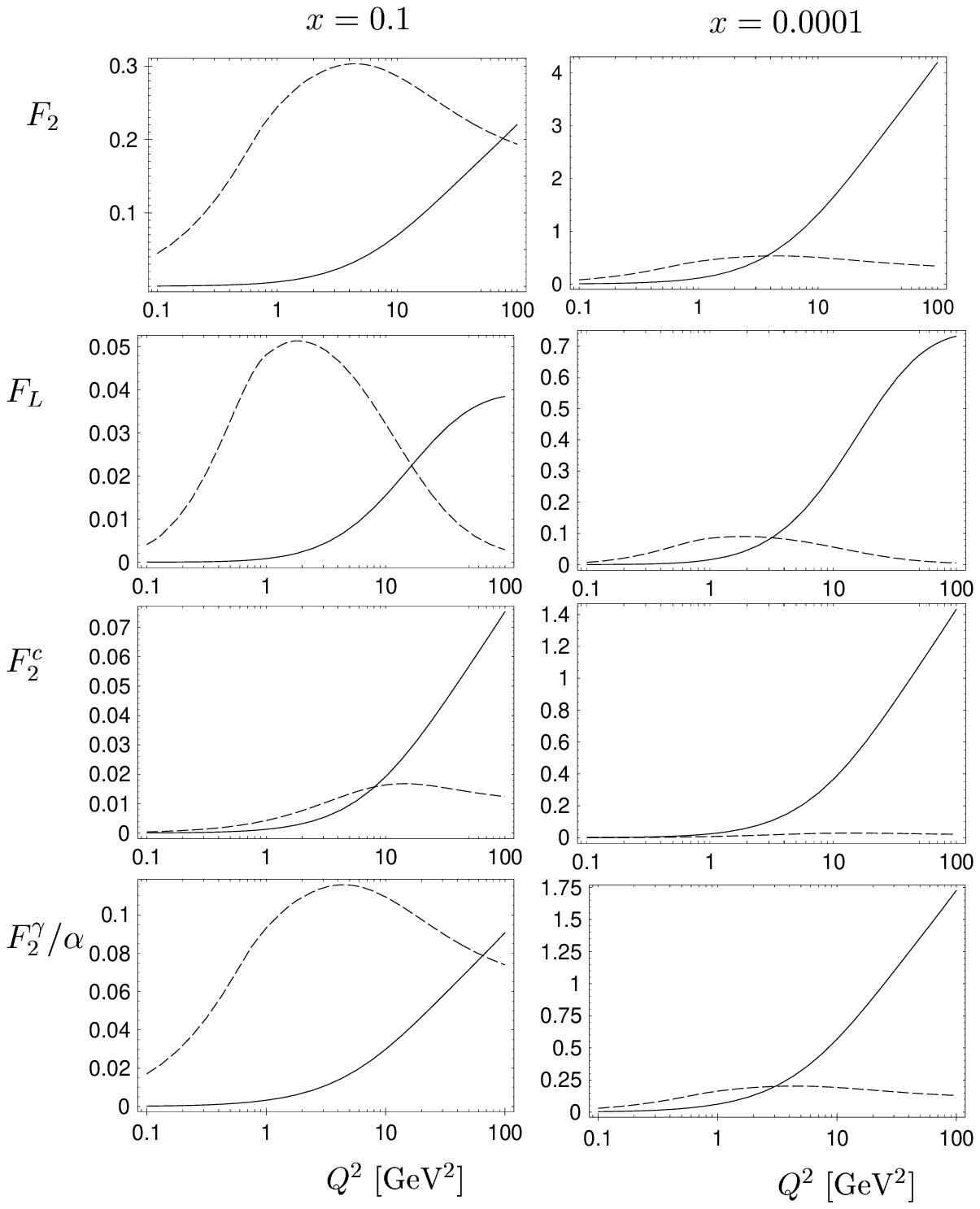}
\end{minipage}
\end{center}
\caption{The soft-pomeron (solid line) and hard-pomeron (dashed line) 
contributions to $F_2$, $F_L$, $F_2^c$ and $F_2^\gamma/\alpha$ at different 
values of $x$.} 
\end{figure}

\section{Conclusions}

This simple dipole-dipole model incorporating a soft and a hard pomeron and 
with only one free parameter provides a remarkably good description of a 
variety of $\gamma^{(*)} p$ and $\gamma^{(*)}\gamma^{(*)}$ data. The only 
reactions for which the model does not provide a successful prediction is 
charm production in $\gamma^{(*)}\gamma$ reactions. In contrast the model 
does provide an excellent description charm production in $\gamma^{(*)} p$ 
reactions, explaining the phenomenological result that $F_2^c$ is completely 
dominated by the hard pomeron. The relative importance of the soft and hard 
pomeron contributions to $F_2$, $F_2^c$, $F_L$ and $F_2^\gamma$ are indicated 
in figure 4. Two features stand out: the dominance of the hard pomeron in 
$F_2^c$ and the comparability of $F_2$ and $F_2^\gamma$.

A consequence of this approach is that the hard pomeron is not restricted to 
hard processes, but is present in soft proceses as well. The real $\gamma
\gamma$ cross section receives a non-negligible and essential hard contribution
due to the point-like coupling of the photon and the resulting singularity of 
the photon wave function at the origin. There is a hard contribution to the 
$\gamma p$ cross section, compatible with although not demanded by experiment. 
There is necessarily a non-zero hard contribution to proton-proton scattering, 
although at energies presently available it is sufficiently small to be within 
the limits of experimental error. However at LHC its presence will be 
observable, although suppressed relative to the simple model by unitarity 
corrections. Broadly similar conclusions have been obtained in related 
analyses, although from a somewhat different standpoint, by Cudell 
{\it et al} \cite{Cud03} and by Donnachie and Landshoff \cite{DL04}.



\begin{thebibliography}{99}
\bibitem{DL}
A. Donnachie and P. V. Landshoff, \PLB~{\bf 437}, 408 (1998)

\bibitem{DD02}
A. Donnachie and H. G. Dosch, \PRD~{\bf 65}, 014019 (2002)

\bibitem{Nac}
O. Nachtmann, Ann. Phys. {\bf 209}, 436 (1991)

\bibitem{Heid1}
A. Kraemer and H. G. Dosch, \PLB~{\bf 272}, 114 (1991)\\
H. G. Dosch, E. Ferreira and A. Kraemer, \PRD~{\bf 50}, 1992 (1994)

\bibitem{Heid2}
H. G. Dosch, T Gousset, G. Kulzinger and H. J. Pirner, \PRD~{\bf 55}, 
2602 (1997)\\
H. G. Dosch, T Gousset and H. J. Pirner, \PRD~{\bf 57}, 1666 (1998)

\bibitem{ZEUS}
M. Derrick {\it et al} (ZEUS Collaboration), \ZPC~{\bf 72}, 3 (1996)\\
J. Breitweg {\it et al} (ZEUS Collaboration), \PLB~{\bf 407}, 432 (1997)

\bibitem{H1} 
S. Aid {\it et al} (H1 Collaboration), \NPB~{\bf 470}, 3 (1996)\\
C. Adloff {\it et al} (H1 Collaboration), \NPB~{\bf 497}, 432 (1997)
  
\bibitem{ZEUSc}
J. Breitweg {\it et al} (ZEUS Collaboration), {\em Eur. Phys. J.} C {\bf 12}, 
35 (2000)

\bibitem{H1c}
C. Adloff {\it et al} (H1 Collaboration), \ZPC~{\bf 72}, 593 (1996)

\bibitem{DLcharm}
A. Donnachie and P. V. Landshoff, \PLB~{\bf 518}, 63 (2001)

\bibitem{DDR00}
A. Donnachie, H. G. Dosch and M. Rueter, {\em Eur. Phys. J.} C {\bf 13}, 
141 (2000) 

\bibitem{Bud75}
V. M. Budnev {\it et al}, {\it Phys.Rep.} {\bf 15C}, 182 (1975)

\bibitem{OPALa}
G. Abbiendi {\it et al} (OPAL Collaboration), {\em Eur. Phys. J.} C {\bf 18}, 
15 (2000)\\
K. Ackerstaff {\it et al} (OPAL Collaboration), \PLB~{\bf 411}, 387 (1997)\\
K. Ackerstaff {\it et al} (OPAL Collaboration), \PLB~{\bf 412}, 225 (1997)  

\bibitem{L3a}
M. Acciari {\it et al} (L3 Collaboration), \PLB~{\bf 436}, 403 (1997)\\
M. Acciari {\it et al} (L3 Collaboration), \PLB~{\bf 447}, 147 (1999)

\bibitem{ALEPHa}
D. Barate {\it et al} (ALEPH Collaboration), \PLB~{\bf 458}, 152 (1999)  

\bibitem{L3}
M. Acciarri {\it et al} (L3 Collaboration), \PLB~{\bf 453}, 83 (1999);
{\it ibid} {\bf 503}, 10 (2001); {\it ibid}

\bibitem{OPAL}
G. Abbiendi {\it et al} (OPAL Collaboration), {\em Eur. Phys. J.} C {\bf 16}, 
579 (2000) 

\bibitem{Cud03} 
J. R. Cudell, E. Martynov, O. Selyugin and A. Lengyel, \PLB~{\bf 587},
78 (2004)

\bibitem{DL04}
A. Donnachie and P.V. Landshoff, \PLB~{\bf 595}, 393 (2004)

\end{thebibliography}
\end{document}